\documentclass[showpacs,preprintnumbers,superscriptaddress,amsmath,preprint,aps,pra]{revtex4-1}

\usepackage{amssymb}
\usepackage{graphicx}
\usepackage{epstopdf}

\usepackage[dvips]{epsfig}
\usepackage{bm}
\usepackage{amsmath}
\usepackage{wrapfig}
\usepackage{slashbox}

\usepackage{float}

\usepackage{color}

\usepackage[citecolor=blue,colorlinks=true]{hyperref}
\usepackage{hyperref}

 \newcommand{\threej}[6]{ \begin{pmatrix}
  #1 & #2 & #3 \\
  #4 & #5 & #6 
 \end{pmatrix}}
\begin{document}
\title{Ultracold Molecule Assembly with Photonic Crystals}
\author{Jes\'{u}s P\'{e}rez-R\'{i}os }

\affiliation{Department of Physics and Astronomy, Purdue University, West Lafayette, Indiana 47907, USA }
\author{May E. Kim }
\affiliation{Department of Physics and Astronomy, Purdue University, West Lafayette, Indiana 47907, USA }

\author{Chen-Lung Hung}
\affiliation{Department of Physics and Astronomy, Purdue University, West Lafayette, Indiana 47907, USA }
\affiliation{Purdue Quantum Center, Purdue University, West Lafayette, Indiana 47907, USA }


\date{\today}

\begin{abstract}
Photoassociation (PA) is a powerful technique to synthesize molecules directly and continuously from cold and ultracold atoms into deeply bound molecular states. In freespace, however, PA efficiency is constrained by the number of spontaneous decay channels linking the initial excited molecular state to a sea of final (meta)stable rovibronic levels. Here, we propose a novel scheme based on molecules strongly coupled to a guided photonic mode in a photonic crystal waveguide that turns PA into a powerful tool for near deterministic formation of ultracold molecules in their ground rovibrational level. Our example shows a potential ground state molecule production efficiency $> 90\%$, and a saturation rate $>10^6$ molecules per second. By combining state-of-the-art cold atomic and molecular physics with nanophotonic engineering, our scheme presents a novel experimental package for trapping, cooling, and optically manipulating ultracold molecules, thus opening up new possibilities in the direction of ultracold chemistry and quantum information.
\end{abstract}
\maketitle

\section{Introduction}
It has been speculated that realizing quantum control on the internal degrees of freedom of cold and ultracold molecules would bring about tremendous applications in quantum computation~\cite{demille2002quantum,Rabl2006}, quantum logic spectroscopy~\cite{Mur2012,Wolf2016}, many-body physics~\cite{baranov2012condensed,JPR2010,Syzranov2016,Eiles2017}, advanced spectroscopic techniques~\cite{Brunken2017,jones2006ultracold}, ultracold chemistry~\cite{quemener2012ultracold,JPR2014,ospelkaus2010controlling,ospelkaus2010quantum}, as well as in the study of  fundamental physics~\cite{carr2009cold,ACME2014}. Unlike with cold atoms, however, the inherent molecular degrees of freedom translate into abundant internal states of molecules, making their production and subsequent manipulation experimentally challenging.

In the synthesis of ultracold molecules, one of the most successful approaches has been the indirect cooling techniques, i.e., assembling them directly from the constituent atoms prepared at cold temperatures ($T<1$~mK), such as in magnetoassociation~\cite{chin2010feshbach} or photoassociation (PA) ~\cite{jones2006ultracold,ulmanis2012ultracold}. Both methods rely on the modification of the scattering properties of colliding atoms through either ramping a magnetic field (magnetoassociation) or addressing the interaction through a laser field (PA). PA, in particular, can be viewed as a light-assisted chemical reaction~\cite{jones2006ultracold,JPR2015,JPR2014}, in which two free atoms absorb a photon resonant with an excited electronic molecular state and form an excited complex that eventually decays back into the continuum or into one of the many ro-vibrational states of the ground electronic state of the complex. To further guide these molecules into stable rovibrational states with sufficient efficiency, coherent quantum state-transfer of an initially dense sample in a \emph{single-pass} via, for example, the stimulated Raman adiabatic passage \cite{ni2008high,danzl2008quantum,lang2008ultracold,danzl2010ultracold,
takekoshi2014ultracold,molony2014creation,park2015ultracold} or a pump-dump method~\cite{sage2005optical} is required. The lack of efficient routes for light-assisted molecular synthesis and closed optical transitions has made it difficult to achieve continuous molecular production and quantum state manipulation. 

In this manuscript, we propose a new paradigm for light-assisted molecular synthesis. Our scheme is based on recent experimental progress in the integration of cold atoms with nanophotonics \cite{vetsch2010optical,goban2012demonstration,thompson2013coupling,hung2013trapped,tiecke2014nanophotonic,goban2014atom,goban2015superradiance}. It has been demonstrated that laser-cooled atoms can be localized in the near field of engineered nanophotonics, namely, photonic crystals \cite{joannopoulos2011photonic}, and achieve strong interaction between single atoms and photons \cite{thompson2013coupling,tiecke2014nanophotonic,goban2014atom,goban2015superradiance}. Our scheme seeks to directly engineer the photonic environment of a molecule trapped near a photonic crystal to create a \emph{nearly-closed} optical transition between its (meta)stable vibronic ground state and an excited state, which is accessible by the free atoms through a single-color PA. This allows photoassociated molecules to be vibronically cooled nearly-deterministically into a stable ground state following spontaneous decay.

Our scheme incorporates a simple Bragg-grating photonic crystal waveguide (BPCW), as shown in Fig.~\ref{fig:concept} (a), that supports tight optical traps for both cold atoms and molecules in the vacuum space near its dielectric surface. Near the BPCW, polarized electromagnetic vacuum fluctuations induce a large light-molecule coupling rate $\Gamma_\mathrm{1D}$ to a waveguide mode that is more than 10 times the molecular radiative decay rate in freespace. This ensures a PA-excited molecule to decay into the rovibronic ground state with over 90\% probability, and even allows subsequent optical manipulation of rotational levels through the strongly coupled waveguide mode. The platform can also efficiently grow a ground state molecular chain, under the consideration of the collective coupling effect, which is further discussed in this manuscript.


\section{The decay rate of an excited state molecule}\label{Sec:decay}

We begin by considering the decay rate of an excited state molecule $|v'\rangle$ to a lower molecular state $|v''\rangle$

\begin{eqnarray}
\label{eq1}
\Gamma_{v''}=\frac{2\mu_0}{\hbar}{\bf d}_{v''}^{\dagger}\cdot \Im[\mathbf{G}(\nu_{v''})]\cdot {\bf d}_{v''},
\end{eqnarray} 

\noindent
where $\mu_0$ is the vacuum permeability, $\hbar$ the Planck constant divided by $2\pi$,  $\nu_{v''}$ the transition frequency between states $v'$ and $v''$, $\mathbf{d}_{v''}$ the dipole matrix element, and $\Im[\mathbf{G}]$ stands for the imaginary part of the classical electric field Green's function from Maxwell's equation for a point dipole source at the location of the molecule. To simplify the notation, in the following discussions we primarily label the vibrational level $v'$($v''$) while omitting other molecular state quantum numbers. To properly work in the molecular frame, Eq.(\ref{eq1}) can be recast in a spherical basis as
\begin{eqnarray}
\label{eq2}
\Gamma_{v''}=\frac{2\mu_0}{\hbar}\sum_{q=0,\pm 1}|d^q_{v''}|^2T_{-q}(\nu_{v''}),
\end{eqnarray}

\noindent
where $T_{q=0,\pm1}$ stands for the spherical component of the Green's tensor responsible for spontaneous decays via $\pi$- and $\sigma^\pm$-transitions, respectively, and $d^q_{v''}$ is the spherical component of the dipole matrix element (\ref{appendixA}).

We separate the total decay rate of an excited state molecule into two contributions, namely,
\begin{eqnarray}
\label{eq:gammatot}
\Gamma_\mathrm{tot}= \Gamma_g + \Gamma'_{ex},
\end{eqnarray}
where
\begin{eqnarray}
\label{eq:gammagandex}
\Gamma_g = \frac{2\mu_0}{\hbar}\sum_q |d_{0}^q|^2 T_{-q}(\nu_0)
\end{eqnarray}
and
\begin{eqnarray}
\Gamma'_{ex} = \frac{2\mu_0}{\hbar} \sum_{v''\neq0} \sum_q |d_{v''}^q|^2 T_{-q}(\nu_{v''})
\end{eqnarray}
are the decay rate to the rovibronic ground state and the total rate to all other vibronic states, respectively. 

Our approach is to engineer a photonic environment to ensure that $\Gamma_g \gg \Gamma'_{ex}$ so that an excited state molecule primarily decays down to the rovibronic ground state. This can be achieved by first identifying an excited molecular state that has significant dipole matrix element for the target molecular ground state $|d_{0}^q|^2>|d_{v''\neq 0}^q|^2$, and then enhance $T_q(\nu_{0}) \gg T_q(\nu_{v''\neq 0})$, for a given $q$, for the radiative decay between the two states through nanophotonics engineering.

\subsection{An example: transition dipole moment (TDM) of Rb$_2$ diatomic molecules}\label{sec_PA}

\begin{figure}[t]
\centering
\includegraphics[width=1\columnwidth,keepaspectratio]{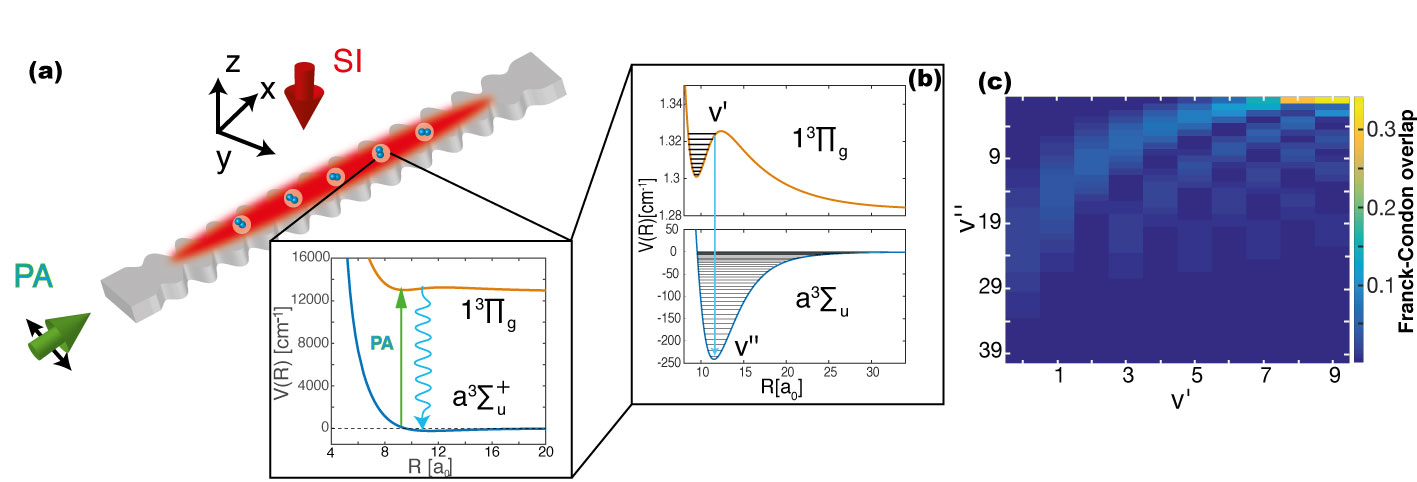}
\caption{Concept of ultracold molecule assembly with a photonic crystal. (a) Schematic of atoms and molecules trapped above the surface of a Bragg-grating photonic crystal waveguide (BPCW). The dipole trap is created by a side illumination beam from freespace (SI, red arrow). A photoassociation beam (PA, green arrow) with a given polarization (black arrow) is launched into the waveguide mode to associate free atoms into ground state molecules. (b) Relevant internuclear potential curves for $^{87}$Rb$_2$ molecular states taken from Ref~\cite{Allouche2012}. Free atoms are photoassociated into the 1$^3\Pi_g$ state followed by enhanced spontaneous decay into metastable $a^3\Sigma_u^+$ state (adopting Hund's case b). (c) Franck-Condon overlap $|\langle v''|v'\rangle|^2$ between initial vibrational state $v'$ and final state $v''$.}
\label{fig:concept}%
\end{figure}

As an example, we consider $^{87}$Rb$_2$ diatomic molecules, whose PA spectroscopy and transition dipole moments are known in the literature~\cite{Allouche2012,bellos2011formation,lang2008ultracold}, although we emphasize that this scheme can be extended to other bi-alkali molecules presenting short-range PA pathways \cite{deiglmayr2008formation,Blasing2016,shimasaki2015production}. Specifically, we calculate the vibrational states of $^{87}$Rb$_2$ associated with the relevant potential energy curves taken from Ref.~\cite{Allouche2012} using the Numerov method \footnote{The calculation is performed in a uniform radial grid from 4.25 a$_0$ up to 450 a$_0$ using 10$^5$ steps for the $a^3\Sigma_{u}^+$ potential, whereas the radial grid for the electronic excited state is taken from 4.25 a$_0$ up to 12.5 a$_0$, with the same number of steps and reaching a convergence error $< $0.1\%}. As schematically represented in Fig.~\ref{fig:concept} (b), we have identified that the vibronic level $v'_\mathrm{PA}=9$ of the 1$^3\Pi_g$ state can be selected for PA synthesis because of its large Franck-Condon overlap, i.e., $|\langle v'_\mathrm{PA}|v''\rangle|^2$ with the ground state $v''=0$ in a$^3\Sigma_u^+$ relative to those of other vibronic levels as shown in Fig.~\ref{fig:concept} (c).

By means of our bound state calculations, we find a large ratio $\eta^{-1} \equiv d^2_0 /\sum_{v''} d^2_{v''}\approx 0.52$, where $d_{v''}\equiv\langle v''||d(R)||v'_{PA}\rangle$ and $d(R)$ stands for the effective molecular TDM for a given electronic transition after the Franck-Condon principle is applied; $d(R)$ depends on the internuclear distance $R$ since it must account for the vibrational degrees of freedom (\ref{appendixA}). In freespace, our calculation suggests already a sizable decay probability into the vibronic ground state $P_g = \Gamma_g/\Gamma_\mathrm{tot} \sim\eta^{-1} \approx 0.52$ . Similar efficient pathway has been experimentally confirmed by Bellos et al.~\cite{bellos2011formation} for $^{85}$Rb$_2$ molecules.

\section{The Bragg-grating photonic crystal waveguide (BPCW) for ground state molecule synthesis}\label{crystal}

Having identified the molecular state for PA, we consider here a simple quasi-one dimensional (1D) Bragg-grating photonic crystal waveguide (BPCW) for enhancing $T_q(\nu_0)$ in Eq.~\ref{eq:gammagandex} and, hence, the decay rate to the ground state. As shown in Fig.~\ref{fig2} (a), the BPCW is formed by a suspended silicon-nitrite (Si$_3$N$_4$) nanobeam with sinusoidal modulations in the dielectric profile that forms the photonic crystal. This simple quasi-1D geometry is expected to offer convenient integration with laser cooling and trapping of cold atoms as well as fabrication with high optical quality \cite{yu2014nanowire}. We note that other nanophotonic designs are also available in the literature, see Refs.~\cite{hung2013trapped,ZangWaveguide16} for example, which in principle can induce even stronger molecular decay rates.

In the following, we discuss how the photonic band structure of the BPCW can be tailored \cite{johnson2001block} to enhance the molecular decay rate. Here, we assume cold atoms and molecules are stabily trapped along the BPCW and postpone the discussion of optical trapping scheme to Section \ref{sec:trap}.

\subsection{ Band structure of the BPCW} 
\begin{figure}[t]
\centering\includegraphics[width=0.8\columnwidth]{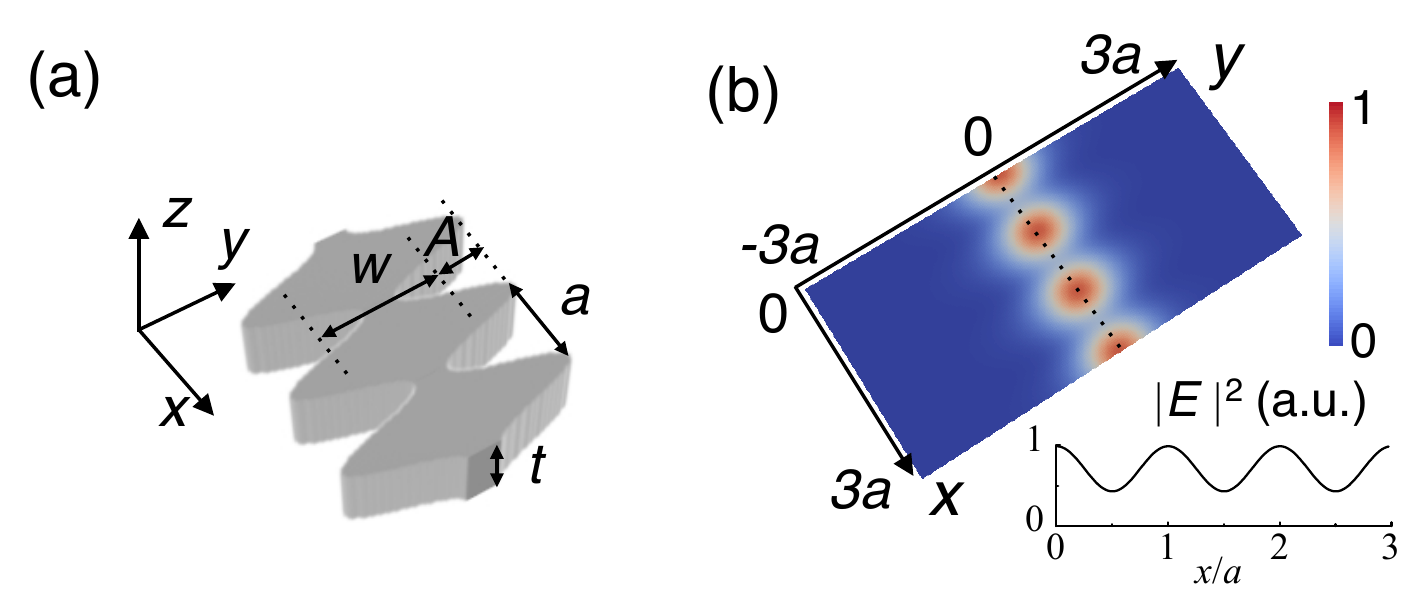}
\caption{The BPCW. (a) Geometry of the BPCW, where $(a,w,A,t)=(291,440,159,247)~$nm are the lattice constant, the nominal width of the waveguide, the sinusoidal modulation amplitude at the outer edges, and the thickness of the BPCW, respectively. (b) Intensity cross-section of the TM mode at the band edge (frequency $\nu_0$), in the $x$-$y$ plane at $z_t=90~$nm above the surface of the BPCW in (a). Inset shows the intensity line cut $|E(x)|^2$ along the $x$-axis through $(y,z)=(0,z_t)~$nm.}
\label{fig2}%
\end{figure}

We have primarily adjusted the BPCW geometry [Fig.~\ref{fig2} (a)] such that the band edge of the fundamental transverse-magnetic (TM) band (electric field primarily polarized along the $z$-axis) is aligned to the transition frequency $\nu_0$ between the ground state of $a^3\Sigma_u^+$ and the chosen PA-excited state at $v'_\mathrm{PA}=9$; see Fig.~\ref{fig3} (a) for the band structure. We expect that the slow light effect near the TM band edge~\cite{baba2008slow,hung2013trapped} can lead to enhanced decay rate into the ground state, discussed in the next section.

Figure~\ref{fig2} (b) shows the intensity cross-section $|E(x,y,z_t)|^2$ of the TM mode at the band edge, where $z_t=90~$nm is the position of the dipole trap center above the surface of the BPCW; see Section~\ref{sec:trap}. We note that the intensity profile modulates periodically along the BPCW, leading to a modulated coupling rate to the waveguide mode $\Gamma_\mathrm{1D}(x)\propto |E(x,0,z_t)|^2$ for trapped molecules along the BPCW as indicated by the dashed line in Fig.~\ref{fig2} (b) \footnote{ The molecules are localized at $(y,z) = (0,z_t)$, and can move along $x$. At this trap location, we find an approximated form $|\Gamma_\mathrm{1D}(x)|^2=\Gamma_\mathrm{1D}\left[0.44+0.56 \cos(\pi x/a)^2\right]$ with the maximum coupling rate $\Gamma_\mathrm{1D}$ occurring at the center of the unit cells, where the width of the dielectric waveguide is the narrowest.}. Since trapped atoms and molecules may freely move along the longitudinal direction of the BPCW (see Sec.~\ref{sec:trap}), the variation of coupling rate along the BPCW can greatly influence the dynamics of molecular array synthesis, discussed in Sec.~\ref{sec:array}.

The band structure of the BPCW can also accommodate all the operations needed for light-assisted molecular synthesis. As seen in Fig.~\ref{fig3} (a), another fundamental transverse-electric (TE) band with electric field primarily polarized along the $y$-axis can be used to guide the PA light and associate free atoms into the exited state molecules. Due to the smallness of the electric field mode area, a moderate power in the guided light can easily saturate the PA transition. We estimate that a total power of 1~mW can lead to $\gtrsim$ 10~kW/cm$^2$ local intensity at the trap location, capable of reaching PA rate $>10^6$ molecules per second~\cite{bohn1999semianalytic}. In addition, an auxiliary probe light can also be launched into the waveguide mode near the edge of the TE-band to detect the presence of cold atoms \cite{goban2014atom} prior to or after PA.

We note that the chosen TM mode (for enhancing spontaneous decay rate) and the TE modes (for photoassociation and probing trapped atoms) are all in the `air' bands, whose field intensities are strongest at the narrowest part of a unit cell (Fig.~\ref{fig2}). This ensures that both the highest PA rates and the largest decay rates to $a^3\Sigma^+_u$ ground state occur at the same location in the unit cells.

\subsection{ Green's tensor analysis of the BPCW} 
\begin{figure}[t]
\centering
\includegraphics[width=0.8\columnwidth]{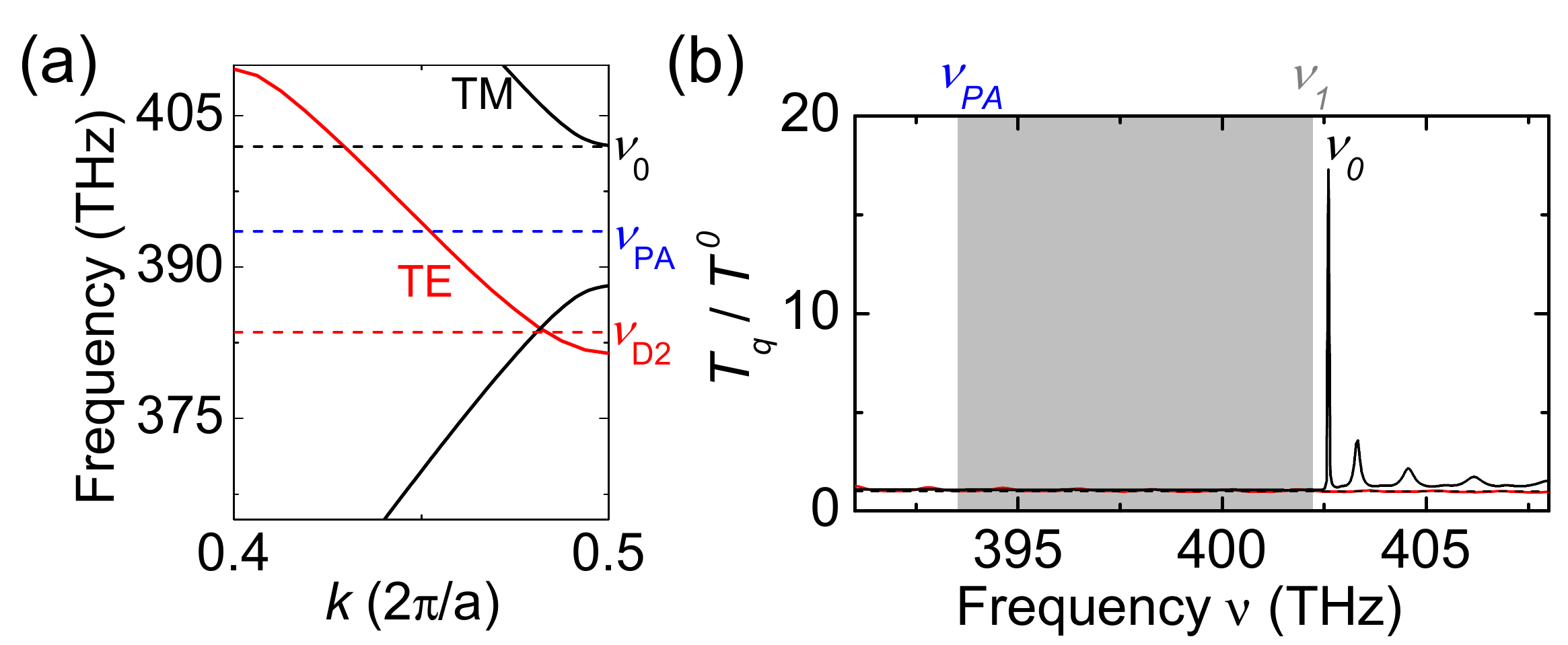}
\caption{Band engineering in the BPCW for ground state molecule assembly. (a) Band structure of the BPCW. Wavenumber $k$ is normalized by $2\pi/a$, where $a=291~$nm is the lattice constant. The edge of the TM `air' band (upper black curve) is aligned to $\nu_0$, the transition frequency to the vibronic ground state (black dashed line). The TE `air' band (red curve) can be used to guide the PA light of frequency $\nu_{PA}$ (blue dashed line). Trapped rubidium atoms can be probed via D2 transition at frequency $\nu_{D2}$ (red dashed line) using either the TE `air' band or the TM `dielectric' band (lower black curve). (b) FDTD calculations of the imaginary part of the Green's tensor, normalized by its freespace value $T^0$ (dashed line). Here, $T_0$ (red curve) and $T_{\pm1}$ (black curve) are responsible for decays via $\Pi$- and $\sigma^\pm$- transitions, respectively; see text. Shaded area marks the frequency range between transitions to the first excited level $v''=1$ ($\nu_1$) and to the dissociation limit ($\nu_\mathrm{PA}$).
}
\label{fig3}%
\end{figure}

To determine the enhanced decay rate to the ground state using Eq.~\ref{eq:gammagandex}, we implement finite-difference time-domain (FDTD) calculations on a sample structure \footnote{The structure considered here consists of 200 unit cells enclosed by additional 3 cells on either end of the BPCW with linear tapering in the modulation amplitude $A$, as defined in Fig.~\ref{fig2} (a). The FDTD Green's tensor calculations presented in Fig.~\ref{fig3} (b) is performed at the SI trap center $(x,y,z)=(0,0,90)~$nm, discussed in Section~\ref{sec:trap}.} to extract the electric field Green's tensor \cite{hung2013trapped}. The Green's tensor is evaluated at the BPCW trap center, $(x,y,z)=(0,0,z_t)$. 

Figure \ref{fig3} (b) shows $(T_0, T_{\pm1})\equiv (\Im[G_{yy}], \frac{1}{2}\Im[ G_{xx}+G_{zz} \pm i (G_{zx} - G_{xz})] )$, where $\Im[G_{\mu \nu}]$ is the imaginary part of the Green's tensor in the Cartesian coordinate and the lab principal axis is aligned to the polarization of the TE-mode [along the $y$-axis in Fig.~\ref{fig2} (a)] since it is used to guide the PA light. 

Near the TM band edge at $\nu=\nu_0$, we observe a sharp peak in $T_{\pm 1}(\nu)$ that is 17 times the free space value $T^0 = 4\pi^2\nu^3/3 c$ for the calculated structure, where $c$ is the speed of light. This peak in $q=\pm1$ components is due to enhanced density of states attributed to the TM band edge mode. Within the frequency range $\nu_\mathrm{PA} \leq \nu \leq \nu_1$, where $\nu_1$ is the transition frequency to the first vibronic excited state $v''=1$ in $a^3\Sigma_u^+$, $T_q(\nu)$ are nearly constant ($\equiv T'$) and equal for all $q=0,\pm1$ due to primarily coupling to radiation channels in freespace; we find $T'\approx T^0$. We define
\begin{eqnarray}
\beta = \frac{T^\mathrm{1D}(\nu_0)}{T'},
\end{eqnarray}
where
\begin{eqnarray}
T^\mathrm{1D}(\nu_0) = T_{\pm1}(\nu_0) - T'
\end{eqnarray}
is the TM-mode contribution. 

The BPCW offers a large ratio $\beta \approx 16$ due to coupling to the TM band edge as seen in Fig.~\ref{fig3} (a). This large $\beta$ will lead to a strong molecular decay rate to the ground state, as discussed in the following section.

\subsection{BPCW enhanced single molecule decay rate and ground state probability}

The above Green's tensor calculation allows us to determine the enhanced molecular decay rate. Specifically, the decay rate to the ground state can now be separated into two terms,
\begin{eqnarray}
\label{eq:gamma1dandg}
\Gamma_g &=& \frac{2\mu_0}{\hbar} \left[ \sum_{q=\pm 1} |d_{0}^q|^2 T^\mathrm{1D}(\nu_0) + \sum_{q=0, \pm 1} |d_{0}^q|^2 T' \right] \nonumber \\
&\equiv& \Gamma_\mathrm{1D} + \Gamma_g',
\end{eqnarray}
where $\Gamma_\mathrm{1D}$ results from the guided mode contribution $T^\mathrm{1D}$, and $\Gamma'_g$ is related to the contribution from all other photonic channels $T'$. 

The ratio between the decay rate to the ground vibronic level and the total decay rate of a single excited state molecule gives the probability $P_g$ for a single molecule to decay into the ground state, which is expressed as
\begin{eqnarray}
\label{gammaratio}
P_g=\frac{\Gamma_g}{\Gamma_\mathrm{tot}} &=& \frac{\Gamma_\mathrm{1D} + \Gamma'_g}{\Gamma_\mathrm{1D}+\Gamma'_g + \Gamma'_{ex}}\nonumber\\
&=& \frac{D\beta +1 }{D\beta  +  \eta},
\end{eqnarray}
where 
$D=\sum_{q=\pm 1} |d_{0}^q|^2/\sum_{q} |d_{0}^q|^2$ is the relative dipole moment that is selected by the BPCW for enhanced decay. 
\noindent
\section{Light-assisted ultracold molecular formation with photonic crystals}\label{results}

In the previous sections, we have introduced the molecular vibronic structure and nanophotonic components for enhancing the ground state PA synthesis. In this section, we elaborate on the specific level scheme relevant to photoassociating and synthesizing ground state $^{87}$Rb$_2$ molecules (Section \ref{sec:levels}), followed by discussions on how the nanophotonic environment described previously plays a pivotal role in achieving near-deterministic formation of rovibronic ground state molecules (Section \ref{sec:ground_state}) and the manipulation of the rotational states for further control (Section \ref{sec:rotation}). Finally, we examine how the collective effect governs the formation of a chain of molecules (Section \ref{sec:array}).

\subsection{Level scheme for PA and ground state molecule synthesis}\label{sec:levels}

\begin{figure}[t]
\centering
\includegraphics[width=0.8\columnwidth]{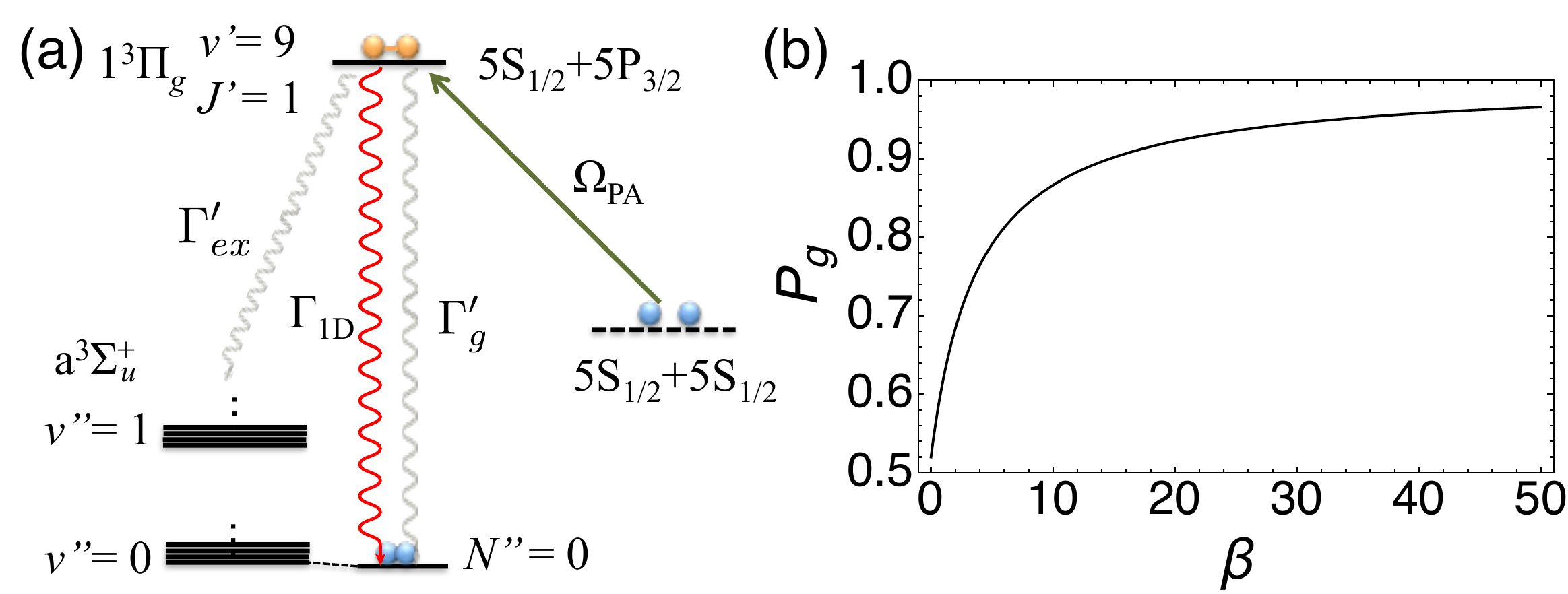}
\caption{Near-deterministic rovibronic ground state molecule synthesis. (a) Detailed level scheme showing incoherent pumping of free atoms at a rate $\Omega_\mathrm{PA}$ to the selected PA state, $v'=9$ and $J'=1$ in 1$^3\Pi_g$, followed by spontaneous decay towards rovibrational levels in the $a^3\Sigma^+_u$. Decay to the ground vibronic level ($v''=0$) is primarily due to a large coupling $\Gamma_\mathrm{1D}$ to the TM band edge (red arrow) and a weak coupling $\Gamma_g'$ to other radiation modes (gray arrow). $\Gamma'_{ex}$ is the total decay rate to all other $v''\neq0$ vibronic levels. Due to selection rules, only the ground rotational level $N''=0$ will be populated. (b) The probability $P_g(\beta)$ for single molecule to decay into the rovibronic ground level.}
\label{fig4}%
\end{figure}

We consider the relevant level scheme in Fig.~\ref{fig4} (a) \footnote{We have ignored the molecular hyperfine structure since typical energy splitting between molecular hyperfine states is $\sim h \times$ 100~MHz, which is an order of magnitude smaller than typical rotational energy splitting $\sim h\times 1$ GHz and is not resolved by the BPCW band structure under consideration.}. While detailed selection rules and dipole matrix element calculations are elaborated in \ref{appendixB}, here we briefly discuss the angular momentum quantum states selected for PA and rovibraonic ground state molecule synthesis. 

In the present case, due to initial ultracold temperature following laser cooling, free $^{87}$Rb atoms will collide in the pure $s$-wave scattering regime through the $a^3\Sigma_u^+$ molecular potential, which is properly described by the Hund's case b. The initial state thus processes even parity~\cite{Blasing2016}, $N=0$ rotational quantum number and $J=1$ total angular momentum quantum number. The PA-excited molecular state in $1^3\Pi_g$, on the other hand, is better described in the Hund's case c with the triad $|J'\Omega' M'\rangle$. Here, $\Omega'$ stands for the projection of the angular momentum in the molecular axis with $M'$ its projection into the lab frame. Considering electric dipole selection rules~\cite{Herzberg,Lefebre,Edmonds}, parity selection rule ($\pm \rightarrow \mp$)~\cite{Herzberg, Lefebre, Blasing2016}, and the relevant TDM (\ref{appendixA}), $J'=1,2$ states can be excited using PA.

As detailed in \ref{appendixB}, we choose to excite the $J'=1$ (and $\Omega'=1$) quantum state using PA light launched in the TE-mode (or polarized along the $y$-axis), with which the $M'=\pm1$ sublevels can be populated. In the subsequent spontaneous decay, parity selection rule dictates that only the rotational ground state $N''=0$ has nonzero dipole matrix element. 

Since the polarization of the TM band edge mode is perpendicular to the TE mode, the BPCW induces $\sigma^\pm$ transitions via enhanced electric field vacuum fluctuations polarized along the $z$-axis, as evident in the peak of $T_\pm(\nu_0)$ in Fig.~\ref{fig3} (b). Following spontaneous decay, the final state should occupy $J''=1$ (and primarily in $M''=0$) and $N''=0$ rotational ground state, and most probably in the $v''=0$ vibronic ground state.

\subsection{Near-deterministic single ground state molecule synthesis}\label{sec:ground_state}

Using Eq.~\ref{eq:gamma1dandg}, we now calculate the total decay rate to the rovibronic ground state in $a^3\Sigma_u^+$, that is $\Gamma_g=\Gamma_\mathrm{1D} +\Gamma'_g = \frac{\mu_0}{\hbar} d^2_0 T'(D \beta + 1)$ with the BPCW. Based on the level scheme discussed above, we calculate that $D=1/2$ (for details, see \ref{appendixB}). With the BPCW, where $\beta\approx16$, we find that the decay rate to the ground state is almost 10 times its freespace value $\Gamma_0 = \frac{\mu_0}{\hbar} d^2_0 T^0$. Decay rates into the other vibrationally excited levels are nearly unchanged. Their sum, $\Gamma'_{ex} = \frac{\mu_0}{\hbar}\sum_{v"\neq 0} d^2_{v"}T' \approx 0.1\Gamma_\mathrm{1D}$, is small compared to $\Gamma_\mathrm{1D}$ due to the large $\beta$ in the BPCW. 

Figure~\ref{fig4} (b) further shows the probability $P_g$ versus a possible range of $\beta$ realized in generic photonic structures. We see that $P_g$ rises significantly from the free space probability $1/\eta\approx 0.52$ towards $P_g=0.91$ for $\beta\approx16$ in our example. For an even larger value of $\beta$, achievable beyond the simple BPCW structure, say Ref. \cite{ZangWaveguide16}, the probability further approaches unity, making the PA synthesis to the ground state molecules deterministic. 

\subsection{Rotational level manipulations}\label{sec:rotation}
Rotational levels can also be manipulated following PA synthesis. For example, we note that $J'=2$ state in $1^3\Pi_g$ couples to both $N''=0,2$ rotation levels in $a^3\Sigma^+_u$ (\ref{appendixB}). Therefore, population transfer between $N''=0$ and $N''=2$ states can be realized either by using two-photon coherent transfer via the $J'=2$ state or simply by optical pumping. The latter is possible because decay rate to either $N''=0, 2$ states can be equally enhanced in a BPCW due to insignificant rotational level spacing $\sim 2~$GHz relative to the vibrational energy splitting $\nu_0-\nu_1\approx 400$~GHz. With large $\beta$ and $\Gamma_\mathrm{1D} \gg \Gamma'_{ex}$, the optical transition takes place in an approximately closed $\Lambda$-system, opening up new possibilities to optically control the rotational levels that were not feasible in freespace. This is particularly important for homonuclear molecules that do not possess permanent electric dipole moments and the transitions between the rotational levels cannot be driven by using radio frequency or microwave fields~\cite{Herzberg,Lefebre,Edmonds}.

\subsection{Molecular array synthesis}\label{sec:array}

Following the discussion on near-deterministic single molecule assembly, we now discuss whether our scheme can be scaled to form an array of molecules with high efficiency. This may be relevant for the study of many-body physics showing intriguing and novel behaviors~\cite{JPR2010,baranov2012condensed}. In particular, we note that when multiple molecules couple to the BPCW, the cooperative effect \cite{chang2012cavity,hood2016atom} will play a significant role in the dynamics of molecular synthesis. 

To simplify the discussion and emphasize the importance of collective effects in the formation of a molecular chain, we consider an effective `three-level' system, consisting of the PA excited state $|e\rangle$, the rovibronic ground state $|g \rangle$, and a state $|s \rangle$, which comprises of all the other excited vibronic levels that are neither pumped nor coupled to the BPCW and incoherently accumulates the undesired decay population. In this scheme, the single molecule decay rate from $|e\rangle$ to $|g\rangle$ is $\Gamma_\mathrm{1D} + \Gamma_g'$, whereas the decay rate to $|s\rangle$ is $\Gamma'_s=\Gamma'_{ex}$ as similarly illustrated in Fig.~\ref{fig4} (a). The evolution of the density matrix of $N_m$ molecules is given by 
\begin{eqnarray}
\label{eq110}
\dot{\rho} = -\frac{i}{\hbar}[H_I, \rho ] + \mathcal{L}_c[\rho] + \mathcal{L}_m[\rho],
\end{eqnarray}
where the dipole-dipole interaction Hamiltonian 
\begin{eqnarray}
H_{I}=\sum_{j,k=1}^{N_m}\frac{\Gamma^{jk}_\mathrm{1D}}{2}\sin (k|x_j-x_k|) \sigma_{ge}^j\sigma_{eg}^k
\end{eqnarray}
and the collective coupling
\begin{eqnarray}
\mathcal{L}_c[\rho]= \sum_{j,k=1}^{N_m} \frac{\Gamma^{jk}_\mathrm{1D}}{2}\cos (k|x_j-x_k|) \left[ 2 \sigma_{ge}^j\rho\sigma_{eg}^k - \left\{\sigma_{eg}^j\sigma_{ge}^k,\rho \right\}\right]
\end{eqnarray}
are both governed by $\Gamma_\mathrm{1D}^{jk}=\sqrt{\Gamma_\mathrm{1D}(x_j)\Gamma_\mathrm{1D}(x_k)}$ that depends on the location of the molecules $x_{j,k}$ along the $x$-axis within a unit cell of the BPCW. In addition, the single molecule decay is governed by the Lindblad term

\begin{eqnarray}
\mathcal{L}_m[\rho]= \sum_{l=s,g}\sum_j \frac{\Gamma'_{l}}{2}\left[ 2 \sigma_{le}^j\rho\sigma_{el}^j - \left\{\sigma_{el}^j\sigma_{le}^j,\rho \right\}\right].
\end{eqnarray}

\begin{figure}[t]
\centering
\includegraphics[width=0.8\columnwidth]{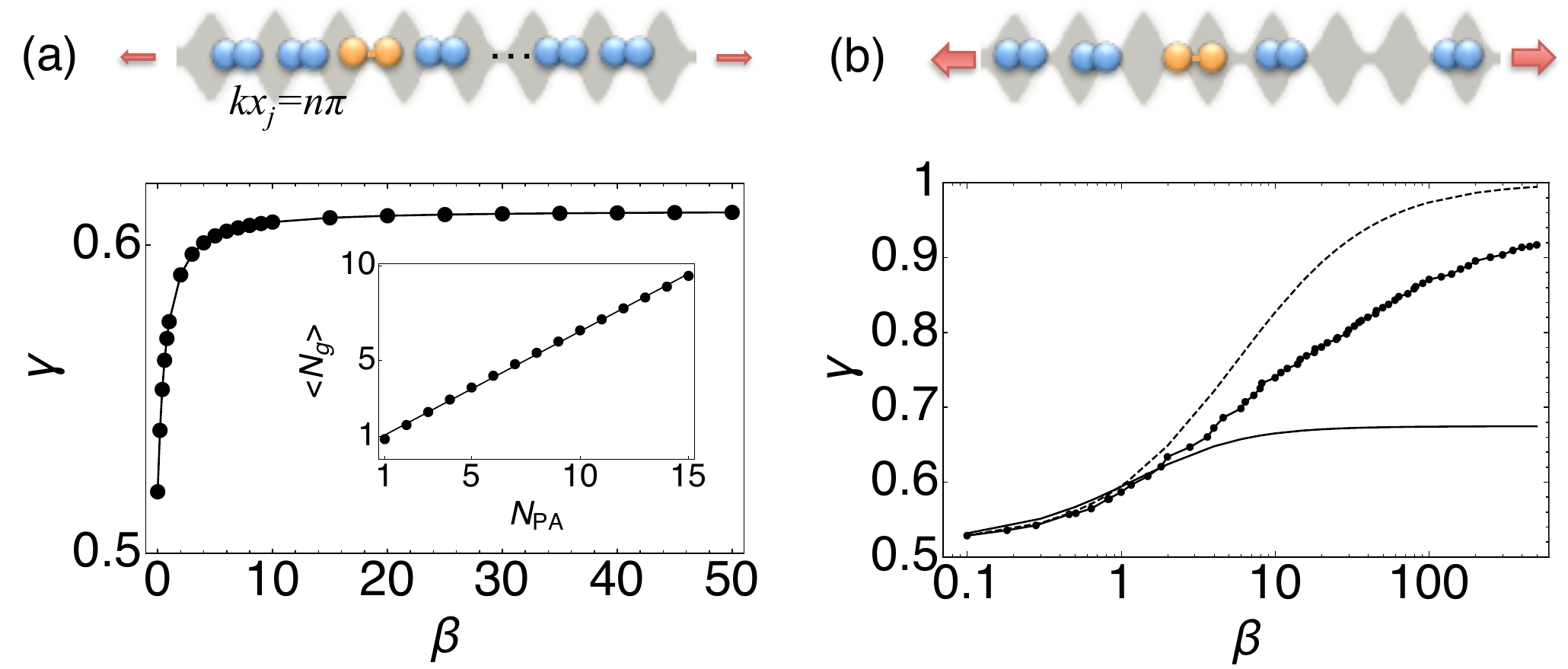}
\caption{Molecular growth dynamics. (a) The averaged ground state molecule growth rate $\gamma$ per PA event, under the localization condition as schematically depicted in the cartoon picture. The linear-growth rate is determined by $N_\mathrm{PA}=15$ consecutively formed PA molecules. Inset shows a sample growth curve at $\beta=10$, where $\langle N_g\rangle$ is the averaged number of ground state molecules. The solid line is a linear fit that determines $\gamma$. (b) The ground state molecule growth rate within an unorganized chain under $N_\mathrm{PA}=6$ consecutive PA events (circles). In contrast to the organized case (solid line), the growth rate is not limited by subradiant coupling and gradually approaches unity in the limit $\beta \gg 1$. The dashed line is the probability $\bar{P}_g$ for $N_m=1$, averaged over random molecule positions along the BPCW.}
\label{fig5}%
\end{figure}

To evaluate the efficiency of molecular array synthesis, we consider PA as a strictly stochastic process, requiring two atoms to randomly come close in space to create an excitation. As a result, we assume only one atom pair can be excited during a single PA event \footnote{This is in contrast to the case of atom-light interaction in photonic crystals \cite{chang2012cavity,hood2016atom}, where an array of atoms can be collectively excited by one photon. We then calculate decay probability using Eq.~(\ref{eq110}), considering only one excited bound pair and $N_m-1$ ground state molecules with no coherence shared among them; as shown in the cartoon pictures in Fig.~\ref{fig5} following decay, $N_m$ increases by 1 if the PA molecule decays into the ground state. Otherwise, it remains unchanged. Beginning from zero ground state molecules, $N_m$ accumulates as the number of PA events $N_\mathrm{PA}$ increases}. In the first case, we assume that all synthesized molecules are localized at the center of the unit cells. That is, $kx_j$ is an integer multiple of $\pi$ where $k\approx \pi/a$ is the wavenumber at the band edge. The localization of molecules leads to a vanishing dipole-dipole interaction in the Hamiltonian, leaving the dynamics governed solely by collective photon emission with constant coupling strength $\Gamma_\mathrm{1D}^{jk}=\Gamma_\mathrm{1D}$. Such decay dynamics is well-known and can be solved under Dicke super- and sub-radiant basis \cite{dicke1954}. 

In Fig.~\ref{fig5} (a), we show the approximated linear growth rate $\gamma$ per PA excitation, determined using up to $N_\mathrm{PA} =15$ consecutive events. Although at low enhancement $\beta$ the ground state molecule growth rate improves beyond the freespace limit, it nonetheless quickly saturates at around $\gamma=0.62$, a number that roughly tracks the overall trend for single molecule decay probability within a chain of $N_\mathrm{PA}$ molecules, as we explain below. 

This undesired inefficiency is due to a growing subradiant contribution in a large molecular chain containing only one PA molecule. Prior to the decay, the molecular chain can be described by a product state $|{g}\rangle_1 \cdots |{e}\rangle_i \cdots |{g}\rangle_{N_m}$, which is a superposition of the Dicke superradiant state with enhanced decay rate $N_m \Gamma_\mathrm{1D}+\Gamma'$, and $N_m-1$ subradiant states of decay rate $\Gamma'$ that do not couple to the BPCW. The probability for ending up with a chain of $N_m$ ground state molecules following single photon decay is expected to be
\begin{eqnarray}
P_{chain} = \frac{1}{N_m}P_g(N_m\beta) + \frac{N_m-1}{N_m} P_g(0),
\end{eqnarray}
approaching $P_g(0)$ at large $N_m$.

To overcome the limitation due to the subradiant effect, in Fig.~\ref{fig5} (b) we consider another case in which molecules are not localized in the unit cells along the BPCW. This likely corresponds to a simple but realistic experiment condition, where cold atoms and molecules can freely along the dipole trap [Fig.~\ref{fig:concept} (a)] between PA events. Due to random phase separations between the molecules and the variation of $\Gamma_\mathrm{1D}(x_j)$ along the $x$-axis within a single unit cell of the BPCW, subradiant effects should be greatly reduced and the excited state molecule may decay approaching the single molecule limit. By numerically \cite{johansson2012qutip} averaging the molecular growth rate of up to $N_\mathrm{PA}=6$ consecutively excited molecules at random positions, we see that the ground state molecule growth rate, indeed, tracks the single molecule limit, even at large $\beta$. A `parasitic' collective coupling effect still lowers the growth rate by $\sim 10$\% at large $\beta$. However, it is no longer limited by subradiance, as in the localized case. We expect this trend to remain valid for larger values of $N_\mathrm{PA}$.

\section{Molecular and atomic trapping potential}\label{sec:trap}
In this section, we discuss how stable dipole traps for cold atoms and molecules can form along a BPCW by way of a side illumination (SI) method \cite{goban2015superradiance,thompson2013coupling}. It has been experimentally realized that cold atoms can be localized in a fixed SI trap near a nanophotonic structure directly using polarization-gradient cooling \cite{goban2015superradiance}. Transportation of cold atoms along a photonic crystal, through steering an SI beam, has also been demonstrated \cite{thompson2013coupling}. 

\begin{figure}[h]
\centering\includegraphics[width=0.8\columnwidth]{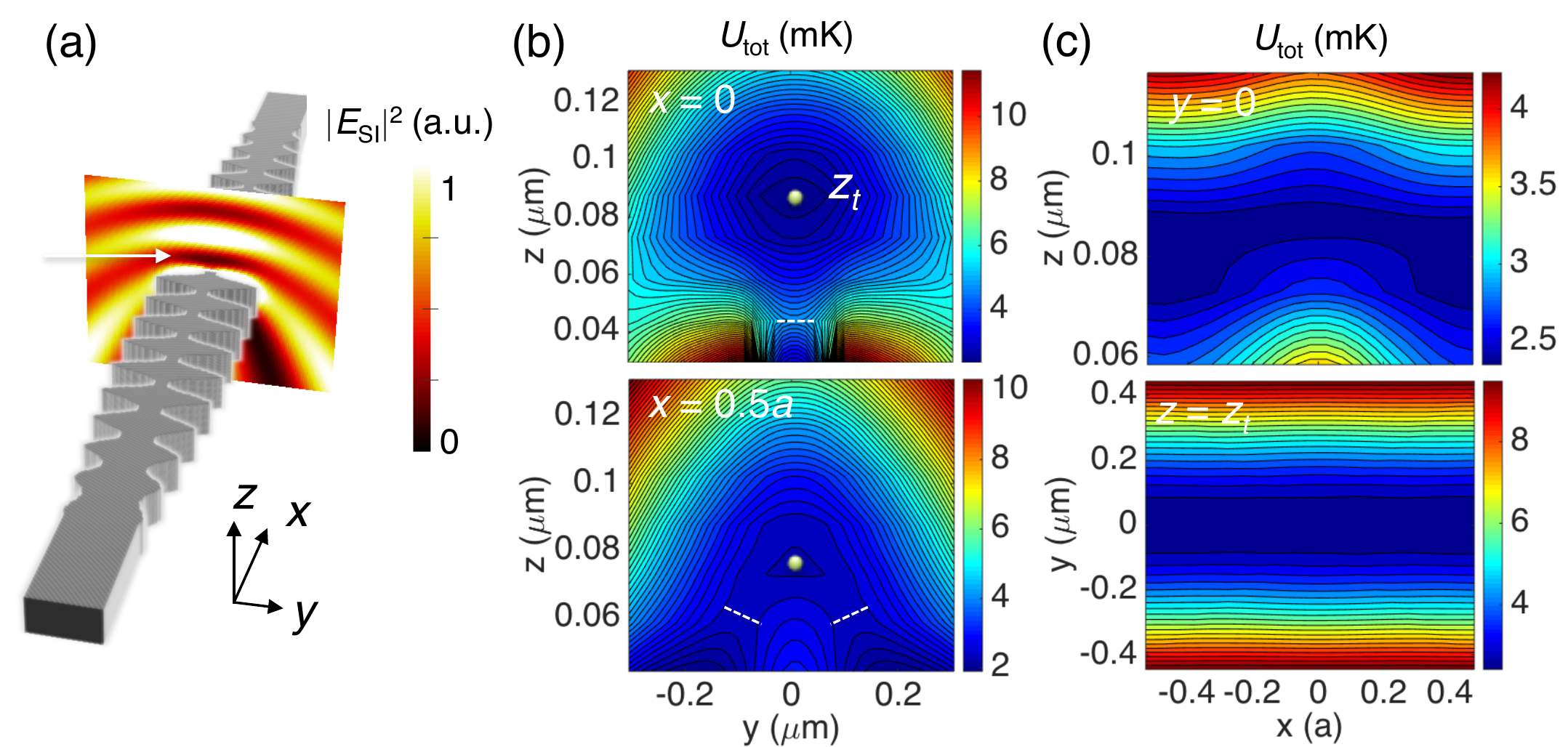}
\caption{ The side illumination dipole trap.
(a) SI intensity cross-section $|E_\mathrm{SI}(0,y,z)|^2$ with $\lambda_b=714~$nm. Gray structure illustrates the BPCW enclosed by the tapering sections. Arrow marks the first interference node around $z_t=90~$nm that forms a blue-detuned SI trap. (b) Cross-sections of the total trap potential $U_\mathrm{tot}$ for Rb atoms at $x=0,0.5a$, respectively. Trap center is marked by green spheres. Dashed lines mark the position where the trap opens, beyond which $U_\mathrm{tot}$ becomes unstable. (c) Cross-sections of $U_\mathrm{tot}$ in the $x$-$z$ plane ($y=0$) and the $x$-$y$ plane  ($z=z_t$), respectively. }
\label{fig6}%
\end{figure}

Figures~1 (a) and \ref{fig6} illustrate the trapping scheme and the trap potential. Specifically, a dipole beam is polarized along the $x$-axis of the BPCW and illuminates from the top of the BPCW. The interference between the SI beam and its reflection off the BPCW can form a stable dipole trap potential near the top surface of the BPCW.

Based on polarizability calculations in Ref.~\cite{Deib2015}, we focus on one illustrative wavelength, $\lambda_b=714$~nm, for a repulsive SI-trapping scheme. The dynamic polarizability $\alpha_m (\lambda_b)$ of the $a^3\Sigma^+_u$ ground state molecule is predominantly real with $\alpha_m(\lambda_b) = -5100$ (a.u.; atomic units), leading to a repulsive dipole force. Meanwhile, for ground state Rb atoms, the polarizability $\alpha_a(\lambda_b)=-1480$ (a.u.) has the same sign but with a smaller magnitude, suggesting that an SI trap for atoms is also stable for the ground state molecules. 

Figure~\ref{fig6} (a) shows the cross-section of the SI intensity $|E_\mathrm{SI}(0,y,z)|^2$, where we assume uniform illumination covering the entire length of the BPCW. The first interference node at $z_t\approx 90~$nm above the surface of the BPCW can form a closed trap for the atoms, $U_\mathrm{SI} = -\frac{1}{4} \alpha_a(\lambda_b)|E_\mathrm{SI}|^2$, and an even stronger trap for the ground state molecule since $\alpha_m(\lambda_b)/\alpha_a(\lambda_b)>3$. We note only a small $\sim3~$\% intensity variation in $|E_\mathrm{SI}(x,0,z_t)|^2$ along the BPCW, but additional trap confinement along the $x$-axis can either be provided by the envelope of the SI beam or by another superimposed dipole potential.

We have also considered surface Casimir-Polder (CP) interactions $U_\mathrm{cp}$. While FDTD calculations \cite{Rodriquez2009theory} can be implemented to determine $U_\mathrm{cp}$ with an arbitrary geometry \cite{hung2013trapped}, here we adopt a simple approximation, $U_\mathrm{cp}(d) = -C_4/d^3(d+\bar{\lambda})$ to estimate the SI beam intensity requirement, where $d$ is the distance to the proximal dielectric surface. For Rb atoms, $C_4\approx 10^{-31}$~J$\cdot \mu$m$^4$ and $2\pi\bar{\lambda}=0.65~\mu$m. For molecules, we assume similar values of $(C_4,\bar{\lambda})$, but detailed calculations will be carried out elsewhere.

Figures~\ref{fig6} (b-c) show the cross-sections of the total potential $U_\mathrm{tot} = U_\text{SI} + U_\text{cp}$ through the SI trap center. To prevent $U_\text{cp}$ from opening an otherwise stable SI trap, in Figs.~\ref{fig6} (b-c) we assume a local SI beam intensity of $130~$mW$/\mu$m$^2$. We determine the trap depth $\Delta U_\text{tot}/k_B\approx$150 $\mu$K, which is limited by the potential difference between the trap center and the trap opening location marked by dashed lines in the lower panel of Fig.~\ref{fig6}~(b). The cross-sections in Fig.~\ref{fig6} (c) further indicate that while transverse confinement is tight, trapped atoms and molecules can freely move along the $x$-axis of the BPCW. Lastly, we note that the simple form of $U_\text{cp}$ leads to an overestimate of the SI power requirement, and an underestimate of the trap depth \cite{hung2013trapped}. 

We note that it is also possible to use an attractive dipole force to simultaneously localize atoms and molecules. An SI beam at a wavelength of $\lambda_r=1064~$nm, for example, can form a stable dipole trap in the first anti-node at $z_t\approx 220~$nm above the surface of the BPCW. Due to the larger trap distance from the surface, the SI intensity requirement is weaker by 10-fold to achieve a similar trap depth. However, coupling rate to the waveguide mode is also $\sim 10$ times smaller than the case with $z_t=90~$nm due to evanescent decay of the TM mode.

\section{Outlook: implications and future work }\label{outlook}

In summary, we present a viable nanophotonic platform designed for trapping, probing and synthesizing ultracold molecules nearly deterministically in their vibronic and  rotational ground state. The proposed design is based on proven technology in quantum optics and cold atom physics~\cite{joannopoulos2011photonic,jones2006ultracold,goban2012demonstration,goban2014atom,goban2015superradiance,ulmanis2012ultracold}. The molecular synthesis involves continuous single-color photoassociation from laser-cooled atoms. There is no need for coherent quantum state transfer that works only in a single shot. Our scheme is inherently different from all present schemes or a proposal involving a high-finesse optical cavity~\cite{Search2004Cavity}. The platform can be scaled to grow a chain of $N_m>6$ molecules with nearly deterministic $>90~$\% efficiency (with $\beta >100$). Growing a large molecular chain with $N_m>100$ is possible, albeit with a reduced efficiency due to subradiant coupling between molecules. Nonetheless, we emphasize that once a large chain of ground state molecules is formed, re-excitation created through the waveguide mode will be largely superradiant \cite{goban2015superradiance,hood2016atom}, making the chain couple even more strongly to the waveguide mode with an enhanced rate $N_m\Gamma_\mathrm{1D}$. The BPCW then serves as a highly efficient light-molecular chain interface for optical manipulation with very high cooperativity, leading to important applications such as molecular quantum memory or quantum gates. Our methodology is extendible to any molecular system showing short range PA pathways such as LiRb~\cite{Blasing2016}, RbCs~\cite{bruzewicz2014continuous} or LiCs~\cite{deiglmayr2008formation}, provided stable trap conditions can be realized. Finally, the assembly and unique tunability of molecular arrays in our approach may help reveal the transition from few-body physics to many-body physics in the presence of dipole-dipole interactions, that is greatly discussed in condensed matter physics. At the same time, these arrays may be employed for the study of state selective ultracold chemical reactions.

\section{Acknowledgements}

We thank Ian Stevenson, Daniel Elliott, and Yong Chen for fruitful discussions. C.-L. H. acknowledges support from the AFOSR-YIP. Funding is provided by the Office of Naval Research (N00014-17-1-2289).

\appendix

\section{Molecular dipole matrix elements}\label{appendixA}
\begin{figure}[t]
\centering\includegraphics[width=0.8\columnwidth]{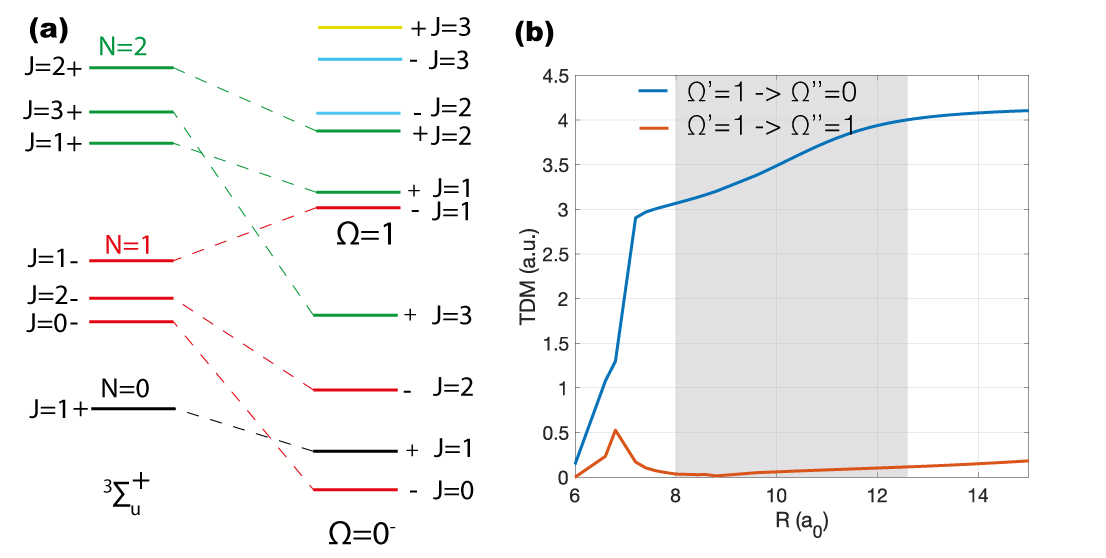}
\caption{ (a) Hund's case b to c relationship for a $^3\Sigma_u^+$ molecular state, adapted from Ref.~\cite{Herzberg}. The symmetry of the state is indicated by $\pm$ equivalent to even/odd symmetry, $N$ stands for the rotational quantum number, and $J$ denotes the total angular momentum quantum number.(b) The radial dependence of the molecular TDMs, i.e., $\langle \phi_{\Omega''} ||d(R)||\phi_{\Omega'}\rangle $ for $1^3\Pi_g-a^3\Sigma^+_u$ transition of Rb$_2$; in particular, $\Omega'=1 \rightarrow \Omega''=0$ transition (blue line) and $\Omega'=1 \rightarrow \Omega''=1$ (red line), taken from Ref~\cite{Allouche2012}. The light gray region represents the intermolecular distances relevant for the vibrational transfer to the ground vibrational state $v''=0$ of the $a^3\Sigma_u^+$ state.}
\label{figTDM}%
\end{figure}

The dipole matrix elements ${\bf d}_{v''}$ may be directly evaluated in atoms, but in the case of molecules, the dipole moment is properly defined in the molecular frame, owing to the inherent dependence of the TDM on the internuclear coordinate $R$. Moreover, the excited electronic states at hand of Rb$_2$ are described in the Hund's case c, where the total angular momentum $J'$, its projection onto the molecular frame $\Omega'$, as well as its projection into the lab frame $M'$ are good quantum numbers~\cite{Herzberg,Lefebre}. The initial scattering state and the final sates following spontaneous decay are, on the other hand, better described in the Hund's case b. However, they can be transformed to the Hund's case c in a one-to-one relationship~\cite{Herzberg} as shown in Fig.~\ref{figTDM}(a). We therefore evaluate the dipole matrix element in the Hund's case c basis. 

Accordingly, one finds ${\bf d}_{v''}=\langle v''J''\Omega''M''|e{\bf r}|v'J'\Omega' M'\rangle$, where the double primed and single primed variables refer to the $a^{3}\Sigma_u^{+}$ and $1^{3}\Pi_g$ electronic states, respectively (see Fig.~\ref{fig:concept}). Therefore, ${\bf d}_{v''}$ is transformed into the molecular frame through 

\begin{eqnarray}
\label{eq4}
d^{q}_{v''}=\sum_{p=\pm 1,0}(-1)^{p-q}d^{p}_{v''}D^{1}_{qp}(\alpha,\beta,\gamma),
\end{eqnarray}

\noindent
where $D^{1}_{qp}(\alpha,\beta,\gamma)$ represents the Wigner D-matrix with $\alpha,\beta$, and $\gamma$ as the Euler angles and $d^q_{v''}$ is the spherical component of $\mathbf{d}_{v''}$. Thus, the effective dipole matrix elements are given by 

\begin{eqnarray}
\label{eq5}
d^{q}_{v''}=\langle v''J'' \Omega'' M''|\sum_{p=\pm 1,0}(-1)^{p-q}d^{p}_{v''}D^{1}_{qp}(\alpha,\beta,\gamma)|v'J' \Omega' M' \rangle,
\end{eqnarray}

\noindent
with~\cite{Varsalovich,Edmonds}  

\begin{eqnarray}
\label{eq6}
|J\Omega M\rangle = \sqrt{\frac{2J+1}{8\pi}}D^{J}_{M\Omega}(\alpha,\beta,\gamma).
\end{eqnarray}

\noindent
For spontaneous decay, we assume degeneracy in the $M''$ and $J''$ sublevels, since we work in low fields and the angular momentum energy splitting and the rotational energy splitting $\sim O$(1)~GHz is insignificant compared to the vibrational splitting $\sim$ 400 GHz between the ground and the first excited state. One can further sum up the contributions from all sublevels, leading to

\begin{eqnarray}
\label{eq7}
|d_{v''}^q|^2 =\sum_{J'',M''}|\langle v'' \Omega'' J''M''|e{\bf r}|v'J'\Omega' M'\rangle |^2 \nonumber \\ 
=  \sum_{J'',M''}(2J''+1)(2J'+1) 
\threej{J''}{1}{J'}{-M''}{q}{M'}^2 \nonumber\\
\left[ \sum_{p}(-1)^p\threej{J''}{1}{J'}{-\Omega''}{p}{\Omega'}\langle v''\phi_{\Omega''} ||d(R)||v'\phi_{\Omega'} \rangle\right]^2, \nonumber \\
\end{eqnarray}
\noindent
where $\phi_{\Omega''}$ and $\phi_{\Omega'}$ label the final and initial electronic states, respectively, and $(..)$ represents the Wigner-3j symbol.

In the present work, the PA state of the molecule is characterized by the $\Omega'=1$ state component which corresponds to the 1$^3\Pi_g$ electronic state of Rb$_2$, labeled as (3)$1_g$ in the Hund's case c notation. However, the $a^3\Sigma_u^+$ state of Rb$_2$ is correlated with two electronic states in the Hund's case c basis: (1)$1_{u}$ and (1)0$_u^{-}$, as shown in  Fig.~\ref{figTDM} (a). The TDMs between the PA state and the two possible final electronic states are shown in Fig.~\ref{figTDM} (b). For the relevant internuclear distances regarding the decay down to the $v''=0$ state (grey region), the TDM for $\Delta \Omega=\Omega''-\Omega'=-1$ is $\sim 100$ times larger than that of $\Delta \Omega=0$. Therefore, $\Delta \Omega=0$ transitions are highly suppressed; in other words, the perpendicular transitions will dominate the decay to the ground vibrational state~\cite{Lefebre}, such that Eq.~(\ref{eq7}) can be further simplified into the form

\begin{eqnarray}
\label{eq8}
|d_{v''}^q|^2=  d_{v''}^2\sum_{J'',M''}(2J''+1)(2J'+1) 
\threej{J''}{1}{J'}{-M''}{q}{M'}^2 \nonumber\\
\times \left[ \sum_{p=\pm 1}(-1)^p\threej{J''}{1}{J'}{-\Omega''}{p}{\Omega'}\right]^2. \nonumber \\
\end{eqnarray}

\noindent
where $d_{v''}^2\equiv \langle v'' ||d(R)||v' \rangle^2$ and the 3j symbol squared involving $\Omega$ is the so-called H\"{o}nl-London rotational line strength factor~\cite{Lefebre}. This equation can be recast as  

\begin{eqnarray}
\label{eq9}
|d_{v''}^q|^2 = d_{v''}^2 \sum_{J'',M''} C(J'',J')\threej{J''}{1}{1}{-M''}{q}{M'}^2, \nonumber \\
\end{eqnarray}

\noindent
where

\begin{eqnarray}\label{eq10}
C(J'',J') = (2J''+1)(2J'+1)\threej{J''}{1}{J'}{0}{-1}{1}^2.
\end{eqnarray}

\noindent
Thus, one finds the selection rule $J''=J'\pm1$ and $J''=J'$, as is customary for electric dipole decay processes. 

\section{Selection rules for rovibronic ground state cooling and optical manipulation of rotational states} \label{appendixB}

The electric dipole selection rules [see Eq.~(\ref{eq10})] impose the allowed transitions for $J'=2$ are to $J''=1,2,3$ and those for $J'=1$ are to $J''=0,1,2$. Taking further into account that the final state should possess even parity, dictated by the parity selection rule $\pm\rightarrow\mp$ ~\cite{Herzberg, Lefebre, Blasing2016}, the allowed transitions for $J'=2$ reduce to $J''=1,3$ while for $J'=1$ only $J''=1$ is possible; see Fig.~\ref{figTDM}(a) for parity of each state. We thus find  
\begin{eqnarray}
\sum_{q} |d_{v''}^q|^2=\frac{1}{2}d_{v''}^2 
\end{eqnarray}
and
\begin{eqnarray}
D = \frac{ \sum_{q=\pm 1} |d_{0}^q|^2}{\sum_{q} |d_{0}^q|^2} = \sum_{J''} \xi_{J''},
\end{eqnarray}
where 
\begin{eqnarray}
\xi_{J''} =  \sum_{M'',q=\pm1}2C(J'',J')\threej{J''}{1}{J'}{-M''}{q}{M'}^2
\end{eqnarray}
indicates the relative population in $J''$ levels after decay. In Table~\ref{SMTable1}, we calculate $\xi_{J''}$ for possible PA states for $J'=1,2$, respectively. Based on Table~\ref{SMTable1}, we summarize the following schemes:
\begin{itemize}
\item \emph{Rovibrational ground state molecule synthesis} Choosing $J'=1$ (and $\Omega'=1$) as the PA state, only $J''=1$ has none zero $\xi_{J''}$. Since $J''=1$ and $\Omega''=0$ (in the Hund's case c) maps to the rotational ground state $N''=0$ in Hund's case b ( see Fig.~\ref{figTDM}), $J'=1$ is the candidate state for synthesizing rovibrational ground state molecules.We note that starting with spin-unpolarized atoms, PA can populate either of the $J'=1, M'=\pm1$ states. The two states have equal $\xi_{J''}$ via coupling to $J''=2$, $M''=0$ state. The selected PA scheme may thus synthesize spin-polarized molecules directly from spin-unpolarized atoms. 

\item \emph{Optical manipulation of rotational levels} On the other hand, $J'=2$ has nonzero dipole matrix element towards $J''=1$ and $J''=3$. The latter correlates with $N''=2$ rotationally excited level. It is thus possible to couple the rotational ground state ($N''=0$) with the excited level ($N''=2$) using, for example, two-photon Raman transition via coupling back to the $J'=2$ state.

\end{itemize}

\begin{table}[h]
\centering
\begin{tabular}{|c|*{4}{c|}}\hline
\backslashbox{$J'(\Omega'=1), M'$~}{$J'' (\Omega''=0)$, parity~}
  & 0,-& 1,+ & 2,- & 3,+ \\\hline
1,$\pm1$ & x & 1/2 & x & x\\\hline
2, $\pm1$  & x & 3/10 & x& 26/105\\\hline
2, 0  & x & 1/5 & x& 8/35\\\hline
\end{tabular}
\caption{$\xi_{J''}$ for the (3)1$_g$ $\rightarrow$ $a^3\Sigma_u^+$ molecule-guided mode coupling. $J', M'$ mark the initial quantum states populated by PA, while $J''$ marks the final state after decay.}\label{SMTable1}
\end{table}  

\newpage
\section*{References}

\bibliographystyle{unsrt}

\bibliography{biblio}

\end{document}